\documentclass[11pt,lettersize]{article}

\usepackage{times}

\usepackage[legalpaper,margin=0.75in]{geometry}

\newenvironment{sciabstract}{%
\begin{quote} \bf}
{\end{quote}}

\usepackage{amsmath}
\usepackage{amssymb}
\usepackage{amsthm}
\usepackage{bm}
\usepackage{graphicx}
\usepackage{xcolor}
\usepackage{cite}
\usepackage{mathtools}  
\usepackage{caption}
\usepackage{subcaption}
\usepackage[T1]{fontenc}
\usepackage[protrusion=true, expansion=true]{microtype}

\DeclareMathOperator*{\argmin}{argmin}
\newcommand{\bigO}[1]{\mathcal{O}\ps{#1}}
\newcommand{\cb}[1]{\left\{#1\right\}}
\newcommand{\complex}{\mathbb{C}}
\newcommand{\est}[1]{\widehat{#1}}

\newcommand{\sq}[1]{\left[#1\right]}

\newcommand{\ip}[2]{\left\langle#1,#2\right\rangle}
\newcommand{\man}{\mathcal{M}}
\newcommand{\norm}[1]{\left\lVert#1\right\rVert}
\newcommand{\ps}[1]{\left(#1\right)}
\newcommand{\abs}[1]{\left\lvert#1\right\rvert}
\newcommand{\reals}{\mathbb{R}}

\newcommand\given[1][]{\:#1\vert\:}
\renewcommand{\v}[1]{\bm{#1}}
\newtheorem{theorem}{Theorem}
\newtheorem*{remark}{Remark}
\newcommand{\normal}{\mathcal{N}}
\newcommand{\complexnormal}{\mathcal{N}_{\complex}}

\newcommand{\Exp}[2][ ]{\mathbb{E}_{#1}\sq{#2}}
\newcommand{\Cov}[2]{\mathrm{Cov}\sq{#1,#2}}
\newcommand{\ft}[1]{\mathcal{F}\cb{#1}}

\graphicspath{{../figs/}}

\title{Stable Filtering for Efficient Dimensionality Reduction of Streaming Manifold Data}

\author
{
  Nicholas P. Bertrand,$^{1+}$%
  Eva Yezerets,$^{2+}$%
  Han Lun Yap,$^{3}$\\
  Adam S. Charles,$^{2\dagger}$%
  Christopher J. Rozell$^{1\dagger\ast}$\\
  \\
  \normalsize{$^{1}$Department of Electrical and Computer Engineering, Georgia Institute of Technology,}\\
  \normalsize{North Ave, Atlanta, GA 30332, USA}\\
  \normalsize{$^{2}$Department of Biomedical Engineering, Johns Hopkins University,}\\
  \normalsize{3400 N. Charles Street, Baltimore, MD 21218, USA}\\
  \normalsize{$^{3}$DSO National Laboratories of Singapore,}\\
  \normalsize{20 Science Park Dr, Singapore, 118230}\\
  \\
  \normalsize{$+$Co-first author},  \normalsize{$\dagger$Co-senior author}
  \\
  \normalsize{$^\ast$Please send correspondences to: crozell@gatech.edu.}
}

\date{}

\begin{document}
\maketitle

\begin{sciabstract}
Many areas in science and engineering now have access to technologies that enable the rapid collection of overwhelming data volumes. While these datasets are vital for understanding phenomena from physical to biological and social systems, the sheer magnitude of the data makes even simple storage, transmission, and basic processing highly challenging. 
To enable efficient and accurate execution of these data processing tasks, we require new dimensionality reduction tools that 1) do not need expensive, time-consuming training, and 2) preserve the underlying geometry of the data that has the information required to understand the measured system. Specifically, the geometry to be preserved is that induced by the fact that in many applications, streaming high-dimensional data evolves on a low-dimensional attractor manifold. Importantly, we may not know the exact structure of this manifold \emph{a priori}.   
To solve these challenges, we present randomized filtering (RF), which leverages a specific instantiation of randomized dimensionality reduction to provably preserve non-linear manifold structure in the embedded space while remaining data-independent and computationally efficient.
In this work we build on the rich theoretical promise of randomized dimensionality reduction to develop RF as a 
real, practical approach.
We introduce novel methods, analysis, and experimental verification to illuminate the practicality of RF in diverse scientific applications, including several simulated and real-data examples that showcase the tangible benefits of RF. 
\end{sciabstract}


\section{Introduction}

Technological developments over the past few decades have vastly accelerated the pace at which data can be collected across a myriad of applications. From new optical microscopy~\cite{song2017volumetric,VaziriMouseData,yu2017two} and cameras~\cite{thomson2022gigapixel} to transcriptomics~\cite{charles2020toward} and social media~\cite{ruths2014social}, scientists and analysts are now faced with a data deluge wherein the bottleneck in transforming raw observations into knowledge has shifted from the acquisition of data to its telemetry, storage, and processing.
For example, in the pursuit of understanding the human brain, neuroscientists are investing massive resources in new neural recording technology with ever higher spatial and temporal resolution~\cite{song2017volumetric,VaziriMouseData,yu2017two,charles2020toward,Jun2017FullyIntegratedSilicon,whang2025fast}.
Should such technologies reach their end goal of simultaneously recording the activity of every neuron in the human brain, the resulting data rate to capture a mere bit per neuron per second would be on the order of 100~Gbps.
Indeed, such an endeavor may not be far beyond the horizon using existing materials and fabrication technologies~\cite{Kleinfeld2019CanOneConcurrently}.
The bandwidth requirements for extracting this data from the system into a digital format will be incredibly high, as even now high dimensional probes are pushing the limits of data read-out~\cite{Jun2017FullyIntegratedSilicon}. 
The ability to reduce the dimensionality of such data in a flexible and efficient manner is paramount to progress in such important data-rich applications.
Moreover, such recording hardware must be heavily power-constrained to achieve non-destructive tissue imaging, further limiting the options for on-board data-compression.

Compression of high throughput streaming data requires techniques that are efficient (fast to compute in power constrained environments), data-independent (do not require training data), and universal (robust to changing data statistics and diverse tasks).
Most compression methods, from classic principal component analysis~\cite{Hotelling1933AnalysisComplexStatisticala} to modern variational autoencoders~\cite{Rezende2014StochasticBackpropagationApproximate, Kingma2013AutoencodingVariationalBayes}, are data-dependent and thus suffer several shortcomings.
First, training data are required to compute the dimensionality reduction map before new data can be compressed. This limits applicability in streaming and online applications which lack such training data. Learning the map also becomes more prohibitive as the data scales, which has prompted continued interest in re-training, fine-tuning, and foundation models. 
Second, such approaches are unsuitable for nonstationary data wherein the underlying statistics drift from the training dataset.
Finally, data acquisition is often intertwined with the discovery process in which the practitioner does not know {\it a priori} which features of the data will be important, making it difficult to determine the appropriate features to preserve. 
Although online PCA~\cite{Boutsidis2015OnlinePrincipalComponents} remedies some of these shortcomings, it incurs a time delay during the online training process, has time complexity which is cubic in its accuracy parameter, and is restricted to a linear subspace model. 
Moreover, the emphasis on maximizing variance explained can be mis-aligned with more analyses, such as classification or regression, that can depend on small but significant variability in the data~\cite{jolliffe1982note}. Capturing only the dimensions of maximum variance is thus often insufficient when compressing data that must later be used for analysis. 

Rather than preserving dimensions of maximum variance explained, we instead aim to preserve the underlying low-dimensional structure present in many high-dimensional datasets. This emphasis is based on the fact that the underlying nonlinear geometry often represents properties of the data that scientists and analysts wish to study. Such structure can arise in many ways, such as the data arising from an underlying dynamical system, or complex physical constraints and correlations due to spatial/compositional properties of the data~\cite{Cunningham2014DimensionalityReductionLargescale,Gallego2018CorticalPopulationActivity,Duncker2021,Langdon2023}. Thus, just as image compression requires compression that preserves perceptual qualities, scientific compression requires compression that preserves these geometrical qualities.  

To provide such a method we present Randomized Filtering (RF), an alternative approach to online data compression based on random projections~\cite{Candes2006NearoptimalSignalRecovery}. RF is computationally efficient, universal, and data-independent, allowing RF to be effective for dimensionality reduction in the context of streaming high-dimensional data. 
Moreover, the rich field of compressive sensing and subsequent work in randomized linear algebra has demonstrated that random projections can preserve data geometry beyond PCA-style compression~\cite{Yap2013StableManifoldEmbeddings,Candes2006NearoptimalSignalRecovery}. Specifically, appropriately designed random projections can preserve data geometry in a variety of structures, including restricted sets, unions of subspaces, manifolds and dynamical systems attractors. 
While traditional approaches considered sets of points that were sparse (i.e., mostly zeros), more general low-dimensional manifold structures can be preserved, as well~\cite{Yap2013StableManifoldEmbeddings}. Past work, however, has been predominantly theoretical and based on projections that were not practically viable in real systems (e.g., random Gaussian projections) or depended on parameters and constants that may not translate well to real-world applications (e.g., assumptions on manifold properties). 
We demonstrate that RF can overcome these challenges to go beyond the theoretical foundations, establishing RF as a practical tool for data compression from dynamical systems measurements across diverse scientific applications. 
We first present RF in a practical algorithm implementation that can be implemented online with low-complexity, show that many real datasets satisfy the appropriate conditions and can be accurately compressed using RF, and present several more detailed case studies in domains including fluid dynamics and neuronal imaging. These examples demonstrate the utility of RF as a general purpose dimensionality reduction tool that fills a critical need as scientific datasets continue to scale.


\begin{figure}[t]
\centering
\includegraphics[height=7cm]{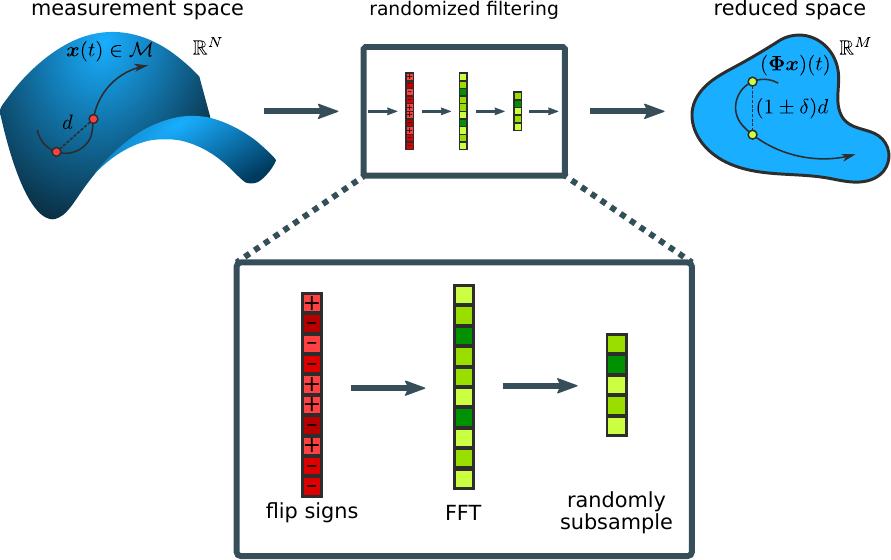}
\caption{
  Illustration of randomized filtering (RF) which consists of three steps:
  1. randomize the signs of the input vector;
  2. compute the FFT of the result;
  3. randomly subsample the Fourier coefficients.
  RF maps points from a $D$ dimensional manifold residing in $\reals^N$ to the reduced space $\reals^M$ where $M<N$.
  Our theory guarantees that for sufficiently large $M$, the mapping is a stable embedding, i.e., pairwise distances are approximately preserved.
  Since the main computational step involves the FFT, the algorithm is of complexity $\bigO{N \log N}$.
}
\label{fig:dimred}
\end{figure}

\section{Results}

\subsection*{RF embeddings maintain data geometry}

In many applications we observe streaming measurements $\v{x}_t\in\reals^N$ for $t=1,2,...$ from a dynamical system that is evolving on a low-dimensional attractor manifold $\v{x}_t\in\man \subset \reals^N$.
The application of RF to $\v{x}_t$ consists of three steps (Fig.~\ref{fig:dimred}): randomly flipping the sign of each measurement $x^{\pm}_{it} = x_{it}$, computing the discrete Fourier transform (DFT) of the sign-flipped result $\v{x}^{\pm}_t$, and finally randomly subsampling the Fourier coefficients~\footnote{The reader may recognize these computational steps as the same ones in the fast Johnson-Lindenstrauss (JL) transform \cite{Ailon2006ApproximateNearestNeighbors}. The important distinction to note is that the theoretical guarantees for RF assume a much more powerful manifold data model which is appropriate for streaming measurements as opposed to the finite point cloud model used in the Fast JL transform.}.
As the algorithmic core is the DFT, the fast Fourier transform (FFT) enables a fast, efficient implementation in only $\bigO{N \log N}$ computations (or even faster if the data is also sparse~\cite{hassanieh2012nearly}). 
Furthermore, the only filter parameters are the binary sign flips and a list of sampled Fourier coefficients. Thus the filter can be generated independently of the data and stored in $\bigO{N}$ memory, which is  more compact than the full linear maps in PCA and far more efficient than highly over-parameterized encoders in autoencoder models (AE).

The benefits of RF are a consequence of the fact that many observations are collected from a $D$-dimensional manifold in an $N$-dimensional ambient space, where $N \gg D$.
Manifolds may be intuitively understood as a generalization of two-dimensional surfaces in three-dimensional space to hypersurfaces embedded in a higher dimension.
Manifold-based modeling has proven effective in describing systems whose state depends nonlinearly on relatively few parameters, e.g., images of objects at different angles~\cite{Donoho2005ImageManifoldsWhich} or neural population activity~\cite{Gallego2018CorticalPopulationActivity}. The core premise underlying RF is that if the dynamical system of interest is evolving on an attractor manifold $\man$ with dimension $D\ll N$, then the system state can be directly compressed to a dimension that scales with $D$ rather than transmitting an amount of data that scales with $N$.

The theoretical underpinnings of RF are based on the notion of \emph{stable embeddings}; i.e., functions that preserve distances between pairs of points in the input space.
Preserving distances is a critical condition to preserving the underlying qualities of the data. Key analyses, including regression, classification, etc., all rely on analyzing the relative distances between points, and any mapping that does not preserve these distances introduces distortions that can affect any subsequent analysis. 
In general, a full isometry, i.e., preserving distances between any two points in a high $N$-dimensional space when mapping to a lower $M$-dimensional space, is impossible. However, given the manifold model, the mapping does not need to preserve all distances and instead needs to only preserve distances between points of interest, i.e., points in the high-dimensional space that are \emph{restricted} on the lower-dimensional manifold. This recommends a more targeted condition: the \emph{restricted isometry property} (RIP).

Formally, we say that a map $\v{\Phi}$ is a stable embedding of $\man$ if for every pair of points $\v{x}, \v{y} \in \man$ we have that the following RIP condition holds
\begin{equation}
1-\delta \leq \frac{\norm{\v{\Phi x} - \v{\Phi y}}_2^2}{\norm{\v{x} - \v{y}}_2^2} \leq 1 + \delta.
\end{equation}
In this condition, the RIP constant $\delta$, a value between 0 and 1, plays an integral role in determining how well distances between points are faithfully maintained by the mapping $\v{\Phi}$. As $\delta\rightarrow 0$, the mapping collapses as points that are distant ($\|\v{x} - \v{y}\|_2^2$ is sizable) can be mapped arbitrarily close together and thereby becoming nearly indistinguishable. Alternatively, as $\delta\rightarrow 1$, the distances between points of interest are exactly maintained, meaning that any function in the high-dimensional space can be exactly replicated in the low-dimensional space. We note that this definition is subtly different than the typical use of RIP, where the manifold is the set of all \emph{sparse} signals (e.g.,~\cite{candes2008introduction,charles2017distributed}), however in this case we describe a much more general case of any low-dimensional manifold. 

Such stable embeddings provide critical robustness to measurement noise that may severely corrupt the measurements during compression.
Bounding the distortion as measured by this generalized RIP provides a guarantee that the map produced by RF is a stable embedding of the input manifold with high probability as long as the number of measurements $M$ satisfies
\begin{equation}
\label{eqn:thm_simple}
  M \geq \frac{K_1}{\delta^2} \ps{D\log\ps{\frac{K_2 N}{\delta}} + \log(K_3)} \log^4(N),
\end{equation}
where $K_i$ are constants that depend on properties of the manifold and the probability of failure\footnote{We note that the number of measurements required may be further reduced beyond the bound given in Theorem~1 by applying appropriate post-processing steps at the filter output \cite{Nelson2014NewConstructionsRIP,Ailon2014FastRIPoptimalTransforms}. However, for simplicity of exposition, we adopt the original procedure described in \cite{Yap2013StableManifoldEmbeddings}.}\footnote{{\bf Theorem 1}
  \textit{
  Let $\man$ be a compact $D$-dimensional Riemannian submanifold of $\reals^N$ with geodesic regularity $R$, volume $V$, and condition number $\tau^{-1}$.
  Let $F \in \complex^{M \times N}$ be a subsampled Fourier matrix whose rows are chosen uniformly from the $N \times N$ DFT matrix, and let $D_\xi$ be a diagonal Rademacher matrix, i.e., the entries along the diagonal of $D_\xi$ are $\pm 1$ with probability $1/2$.
  If
  \begin{equation}
  \label{eqn:thm_condition}
  M
  \geq
  \frac{C_1}{\delta_\man^2}
  \ps{D \log \ps{\frac{RN}{\tau\delta_\man}}
       + \log \ps{\frac{V}{\rho}}
  }
  \log^4(N) \log\ps{\rho^{-1}},
  \end{equation}
  then with probability greater than $1 - C_2 \rho$, $\Phi = FD_\xi$ stably embeds $\man$ with isometry constant $\delta_\man$.
  Universal constants that do not vary with any other quantities in the theorem are denoted by $C_1$ and $C_2$.
  }}~\cite{Yap2013StableManifoldEmbeddings}.
Note that the number of required measurements grows linearly in $D$ (which may be understood intuitively as the amount of ``information'' encoded by the manifold), and only logarithmically in the ambient dimension $N$. Note again the key dependence on $\delta$: as the mapping becomes more faithful, the denominator approaches 1 and the lower bound on the $M$ reduces.

\begin{figure}[t]
\centering
\includegraphics{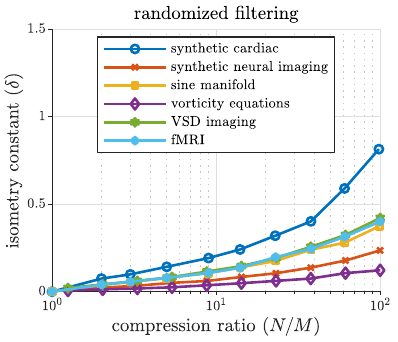}
\includegraphics{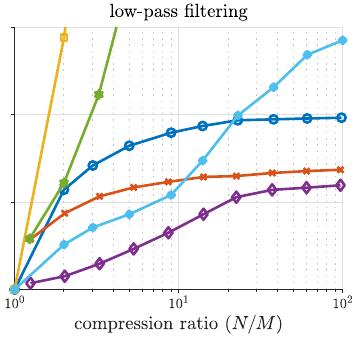}
\caption{
  The isometry constant $\delta$ quantifies how well an embedding preserves the geometry of the input space.
  A value of $\delta = 0$ corresponds to a perfect embedding where distances between all pairs of input points are equal to corresponding distances in the reduced space.
  Shown are estimates of the isometry constant for several synthetic datasets (cardiac model \cite{Elshrif2014QuantitativeComparisonBehavior}, the neural imaging data used in Fig.~\ref{fig:ca_sim}, the sine manifold \cite{Eftekhari2015NewAnalysisManifold}, and solutions to the vorticity equations as in Fig.~\ref{fig:vorticity}) as well as real datasets (voltage sensitive dye imaging from rodent experiments \cite{Zheng2015AdaptiveShapingCortical} and functional magnetic resonance imaging \cite{Nooner2012NKIRocklandSampleModel}).
  Lower isometry constants may be achieved with far fewer measurements with RF (left) compared to LPF (right) which lacks similar stability guarantees.
}
\label{fig:rip}
\end{figure}

The use of random projections for dimensionality reduction stems from the seminal work of
Johnson and Lindenstrauss~\cite{Johnson1984ExtensionsLipschitzMappings}, who showed the existence of a random projection operator that stably embeds arbitrary point clouds. 
Ensuing work expanded this theory to random embeddings that can be applied efficiently~\cite{Ailon2013AlmostOptimalUnrestricted} as well as to proving embedding results for more general manifolds~\cite{Baraniuk2009RandomProjectionsSmooth}.
Our theory unifies these results, providing an efficient embedding based on random projections that also generalizes to manifold data models.
For streaming scientific measurements where the high-dimensional data is governed by an underlying low-dimensional dynamical system, combining the manifold model generalizations with efficient computation is critical for model-free online compression.

Although our theoretical results guarantee that RF is a stable embedding when condition~\eqref{eqn:thm_simple} is satisfied, we seldom have access to the manifold parameters involved in this expression.
It may thus be unclear how many measurements are required for practical application.
Given representative data, we can empirically explore the isometry constant\footnote{For a given dataset, pairwise distances are exhaustively computed in the input and output space and the most severely distorted distance is used to compute the isometry constant $\delta$.  The value of $\delta$ is averaged over instantiations of the randomized filtering parameters.} $\delta$ as a function of the compression ratio $N/M$.
Remarkably, RF produces low estimates of $\delta$ across a wide range of synthetic and real datasets despite using no prior knowledge of the particular details for each data type (Fig.~\ref{fig:rip} and supplemental text).
As a point of comparison, low-pass filtering (LPF)---a dimensionality reduction method with our desired properties of efficiency, universality and data-independence---has higher isometry constants and therefore cannot preserve the data geometry critical for ensuing analysis.

To further assess the applicability of RF in practice, we explore how its geometry preservation translates to real-world data processing tasks: neural activity inference from calcium imaging and flow classification from vorticity measurements.

\subsection*{RF embeddings preserve neural activity in calcium imaging of neurons}
One rapidly growing area where fast, online compression will be critical to stemming the data deluge in the near future is in high-density neural recordings such as optical functional microscopy~\cite{denk1990two,Smetters1999DetectingActionPotentials,whang2025fast} and high-density electrode designs~\cite{Jun2017FullyIntegratedSilicon}.
While necessary to capture the many nuances of neural activity, the extremely high resolution (and therefore bitrate) of these datasets limits the ability to efficiently process them or to potentially transmit the data wirelessly from freely behaving subjects with chronic implantations.
Here we consider one such recording technology---calcium imaging---which is both a staple neuroimaging technique and rapidly being expanded to larger neural volumes with higher fidelity~\cite{yu2017two,song2017volumetric,kong2015continuous}.

\begin{figure}[t]
\centering
\includegraphics[width=0.9\textwidth]{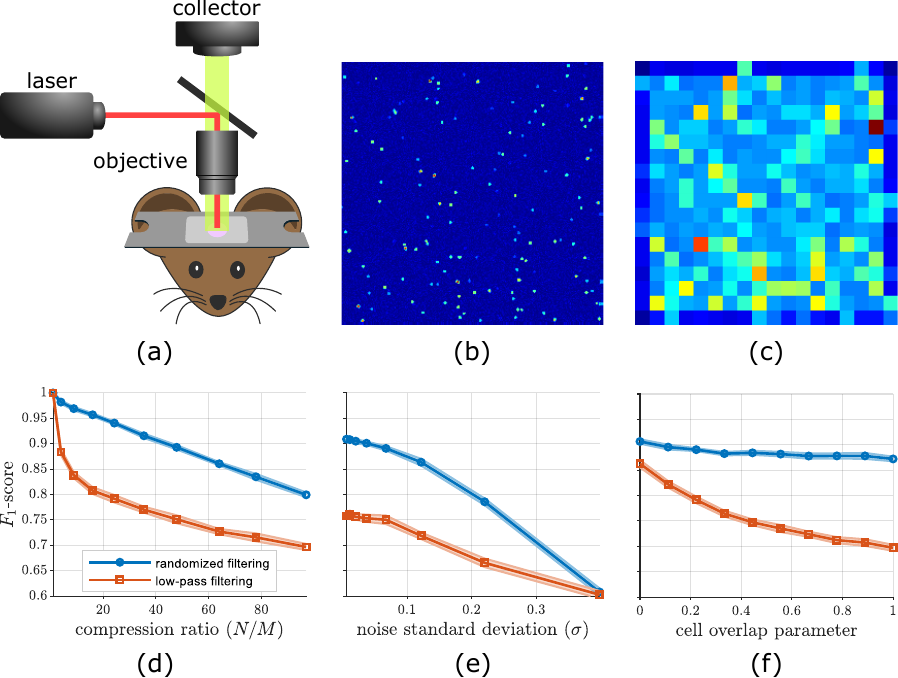}
\caption{
  {\bf (a) } Calcium imaging uses florescent calcium indicators to measure activity in the brain.
  {\bf (b) } Example frame of simulated  calcium imaging data.
  {\bf (c) } In LPF, each frame is blurred and downsampled by $15$X in each spatial dimension as a crude form of dimensionality reduction.
  The fine details of the image are lost, especially in areas containing overlapping cells.
  {\bf (d-f) } Performance comparison for event detection in synthetic calcium imaging data.
  Events are estimated by thresholding the output of Equation~\eqref{eqn:ip}, and the $F_1$-score is used as the performance metric (higher is better).
  {\bf (d)} RF produces favorable results even after heavy compression.
  {\bf (e) } RF outperforms low-pass filtering (LPF) up to moderate levels of noise.
  {\bf (f) } As the number of overlapping cells increases, LPF cannot distinguish activity between nearby or overlapping cells. In contrast, RF can separate activity between cells with significant levels of overlap.
}
\label{fig:ca_sim}
\end{figure}

In calcium imaging (Fig.~\ref{fig:ca_sim}a), fluorescent proteins~\cite{akerboom2012optimization,zhang2023fast} are introduced into neurons (or other cells, e.g., astrocytes~\cite{khakh2015astrocyte}) of interest. Once expressed in the cells, the proteins remain inert until they bond with calcium ions. Once bonded, the proteins can be imaged with optical microscopy techniques (e.g., multi-photon, lightsheet, or widefield technologies). The result is a series of fluorescence images (or volumes) that must be processed to extract the activity time traces of individual fluorescing units (e.g., neurons)~\cite{benisty2022review}.  While the overall bitrate of these technologies can be extremely high, the underlying activity of interest is actually much lower-dimensional. In fact, simple models, such as row-rank, have been highly effective at post-hoc denoising~\cite{pnevmatikakis2016simultaneous}, source extraction~\cite{mukamel2009automated}, and are even being built into the imaging pipeline for fast image acquisition~\cite{whang2025fast}. Here we instead ask if the data can be compressed online as it comes in with RF, and if we can leverage the geometry preservation of RF to extract the activity of functional units directly from the compressed data.

To test the applicability of RF to such datasets, we generate synthetic calcium imaging data with the goal of recovering neural activity directly from reduced measurements~\footnote{Data are generated by first placing $N_c$ circular neuron cell profiles in the $W \times H$ pixel scene.
  Each neuron may overlap with the previously generated one with a fixed probability.
  Temporal dynamics are generated by convolving a Poisson spiking process over $N_t$ time samples which we subsequently convolve with a Gaussian kernel intended to approximate the temporal profile seen in calcium imaging.} (Fig.~\ref{fig:ca_sim}b).
Simulated data affords full control over the data characteristics which permits the assessment of RF in specific parameter regimes.
Denote by $\v{A} \in \reals^{N \times N_c}$ the matrix whose columns correspond to vectorized cell profiles, i.e., the nonzero values in each column are the pixels in space corresponding to one cell. Similarly, let $\v{S} \in \reals^{N_t \times N_c}$ be the matrix whose columns contain the temporal profiles of each cell, i.e., the changes in fluorescence due to changing calcium concentrations. Given these definitions we can model the data matrix as
\begin{gather}
    \v{D} = \v{AS}^T + \sigma \v{\epsilon},
\end{gather}
where the $t^{th}$ column of $\v{D}$ represents the measured fluorescence frame at time $t$, and $\sigma \v{\epsilon}$ represents additive white Gaussian noise with standard deviation $\sigma$.
The compressed data can thus be represented by applying the RF operator $\v{\Phi}$ to each frame independently, a fast operation that can be done as the frames are acquired. The lower-dimensional data is thus $\widetilde{\v{D}} = \v{\Phi}\v{D} = \v{\Phi}\v{AS}^T = \widetilde{\v{A}}\v{S}^T$.

In the absence of noise and cell overlap, the time series for cell $i$ (denoted by $\v{t}_i$) may be recovered via the inner product of the data matrix with its spatial cell profile~\footnote{In the presence of Gaussian noise, this can be shown to be the maximum likelihood estimator.
  Cell overlap may be considered as an additional source of noise.}:
\begin{equation}
\label{eqn:param_est_original}
\est{\v{s}}_i = \v{D}^T \v{a}_i/\norm{\v{a}_i}_2^2,
\end{equation}
where $\v{a}_i$ is the spatial profile of the $i^{th}$ cell. 
Although we do not have theoretical guarantees for complete recovery of the original data, we can still estimate quantities of interest in the reduced space \cite{Davenport2010CompressiveSP}.
In particular, under mild assumptions, stable embeddings also preserve inner products, and we can approximate the uncompressed estimate via
\begin{equation}
\label{eqn:ip}
\est{\v{s}}_i = \Re\cb{\v{(\Phi D)}^T \v{\Phi a}_i}/\norm{\v{\Phi a}_i}_2^2 = \Re\cb{ \widetilde{\v{D}}^T \widetilde{\v{a}}_i}/\norm{\widetilde{\v{a}_i}}_2^2,
\end{equation}
where $\v{\Phi}$ is the RF operator (supplemental text, \footnote{Since the measurements are complex-valued, we need to take the real component of the inner product shown.
  The full derivation for this estimator is provided in the supplemental materials.
  Alternatively, one may consider an efficient real-valued RIP matrix and apply the approach in \cite{Yap2013StableManifoldEmbeddings} to produce a similar guarantee as Theorem 1.
  For example, operators based on the subsampled discrete cosine transform are real-valued, efficient, and RIP optimal, although their construction is slightly more complicated than the one presented here \cite{Ailon2014FastRIPoptimalTransforms}.}).
This estimator in the reduced space is an inner product similar to the one in the original space, but with the dimensionality reduction operator applied to both the data and cell profiles.
Using RF with Equation~\eqref{eqn:ip}, we recovered the time traces for each of the simulated sources pre- and post- compression. As a point of comparison, we applied the same procedure using the main alternative method that is on-line, parametric, and does not require learning: low-pass filtering (LPF) (Fig.~\ref{fig:ca_sim}c). 

To test the quality of the recovered traces, we compared the ability to recover the events in the time traces. These events, also called `transients', are sharp rises in the fluorescence, followed by slow decays back to baseline, that represent the action potentials or burst of action potentials that constitute neural firing. In the case of RF compression, we were able to recover the events accurately using a fraction of the original measurements (Fig.~\ref{fig:ca_sim}). To quantify this we computed the F1 score, defined as
\begin{gather}
    F_1 = \frac{2}{\frac{1}{Precision}+\frac{1}{Recall}},
\end{gather}
where the \textit{precision} represents the number of true positives as a fraction of all detections, and the \textit{recall} represents the number of true positives as a fraction of all actual events. 
The $F_1$-score of the detections stayed above 0.9 for compression ratios of up to 40X (Fig.~\ref{fig:ca_sim}d). LPF performed far worse, quickly dipping below an $F_1$ of 0.9 for compression ratios $<5$X (Fig.~\ref{fig:ca_sim}d). 
Similarly, RF performed much better in the $F_1$ score for detections as the detection problem was complicated by either increasing the noise levels (Fig.~\ref{fig:ca_sim}e), or by increasing the number of spatially overlapping  neurons (Fig.~\ref{fig:ca_sim}f).

\subsection*{The geometry of neural dynamics are maintained in RF embeddings}

\begin{figure}[t]
\centering
\includegraphics[width=0.95\textwidth]{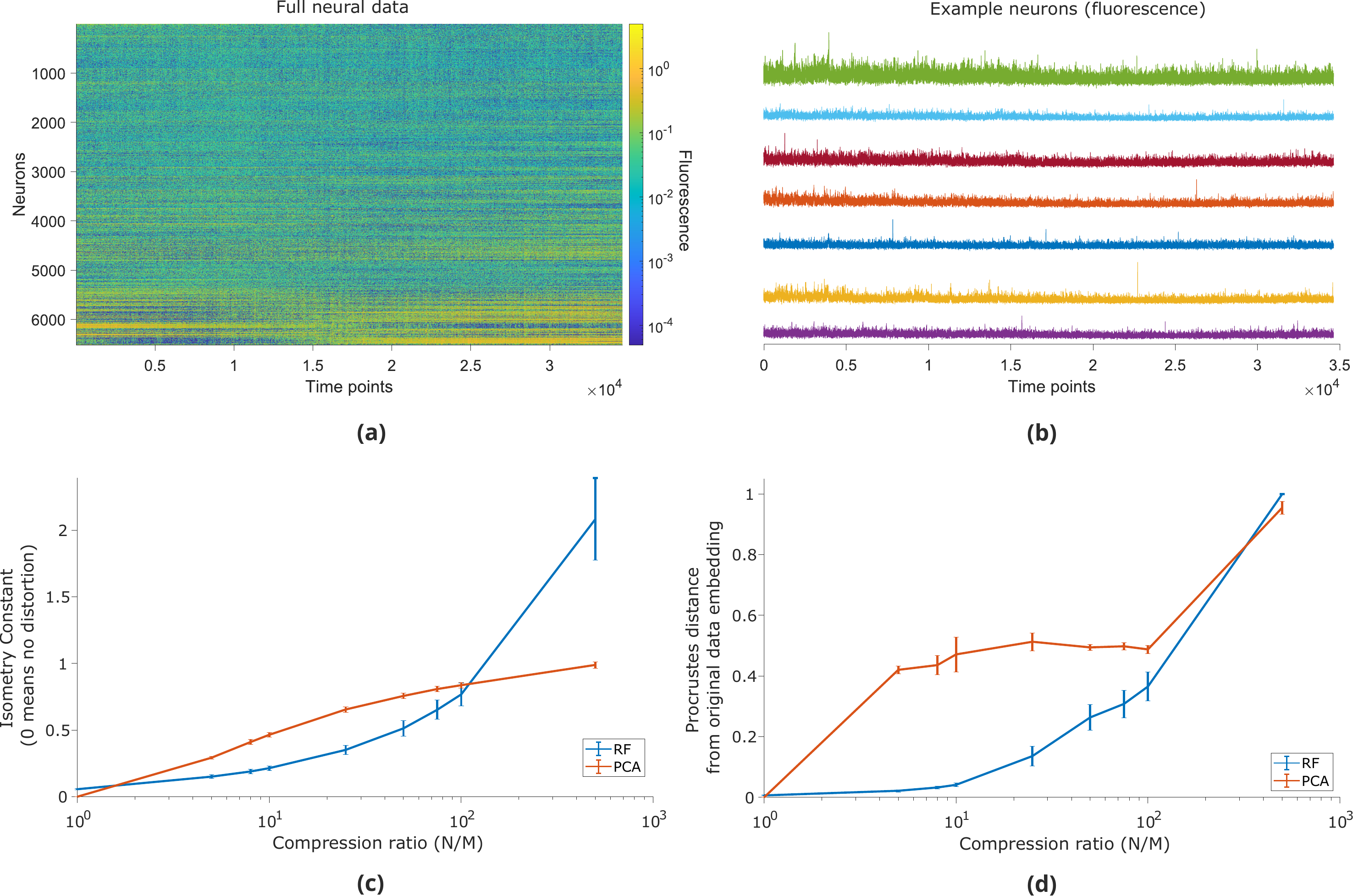}
\caption{
{\bf (a) } Mouse neural fluorescence data, 34600 time points recorded at 9.6~Hz from 6519 neurons~\cite{Manley2024,VaziriMouseData}.
{\bf (b) } Example neurons selected from the data.
{\bf (c) } RF preserves pairwise distances better than PCA for mouse brain data across a range of compression ratios, as measured by the isometry constant $\delta$. 
PCA results are averaged across 100 samples (without replacement, 80 percent of the data, error bars: standard deviation). Randomized filtering results are averaged across 10 samples, each run through 10 random seeds of randomized filtering.
{\bf (d) } RF compression better preserves overall embedded manifold shapes of the data than PCA compression. We embedded each of the 34600 time points from the original data (6519 dimensions), and the RF and PCA compressed data into 4 embedding dimensions using LLE separately for each data matrix. We compared the manifold shapes in the 4-D space using the Procrustes distance between the data embedding and 4-D embeddings of RF- and PCA-compressed data (compression ratios on the x-axis). A Procrustes distance of 0 indicates no difference in shape between the two embeddings. RF preserves the overall 4-D manifold shape better than PCA at compressions up to approximately 100x (65 dimensions).
}
\label{fig:VaziriMouse}
\end{figure}

While neural imaging represents one significant form of big data in neuroscience, cutting edge techniques that record tens-of-thousands of neurons can produce large datasets even after the neural activity traces are extracted. These datasets themselves have complex internal geometry, often termed ``neural manifolds''. While many models and methods have been created to try and estimate such underlying manifolds from data~\cite{Roweis2000LLE}, we argue that the presence of this manifold means that the data is inherently compressible, and that the manifold can be stored and recovered just as well from the compressed data as the neuron-space data.

To validate the manifold-preserving capabilities of RF, we considered real neural calcium imaging data from a mouse brain (Fig.~\ref{fig:VaziriMouse}a,b)~\cite{Manley2024} consisting of approximately an hour-long recording (34,600 timepoints at 9.6~Hz) of 6,519 individual neurons (Fig.~\ref{fig:VaziriMouse}a,b). Denoting the original data  $\v{S}$ (neurons-by-time), we observed that compressing the data with RF, $\widetilde{\v{S}}_{RF} = \v{\Phi}\v{S}$, achieved a lower isometry constant $\delta$, even when compared to PCA compression, despite PCA being data-driven. Specifically, significant improvements were noticeable between compression ratios of 5 and 100 (Fig.~\ref{fig:VaziriMouse}c). Above 100X compression, both methods had isometry constants close to 1.

While the isometry constant provides one measure of distance preservation, we aimed to explicitly test the manifold-preserving capabilities of RF through the lens of a subsequent manifold learning method: Locally Linear Embedding (LLE). LLE is a classical method to identify low-dimensional manifold coordinates given higher-dimensional data. We wanted to ensure that the learned parametrization through LLE was preserved in the RF compression, and therefore ran LLE on the raw data as well as the RF- and PCA-compressed data (Fig.~\ref{fig:VaziriMouse}d). LLE produced for each of these data a 4-D representation of each time-point. To measure how closely the LLE outputs for each of the compressed data versions were to the LLE output of the original data, we computed the Procrustes distance, which measures the difference between two point clouds up to an affine transformation (a rotation and a scaling). Mathematically, if $\v{L}$, $\v{L}_{RF}$, and $\v{L}_{PCA}$ are the LLE embeddings of the original data time traces $\v{S}$, the RF compressed data $\widetilde{\v{S}}_{RF}$, and the PCA compressed data $\widetilde{\v{S}}_{PCA}$, respectively, the Procrustes distance is
\begin{gather}
    d_{proc}(\v{L}, \v{L}_{PCA\text{ \textit{or} }RF}) = \min_{\v{G}} \left\| \v{L} - \v{G}\v{L}_{PCA\text{ \textit{or} }RF}\right\|_F^2,
\end{gather}
where $\v{G}$ is the matrix that defines the affine transformation.
Lower Procrustes distances between the pre- and post-compression manifolds indicate a higher degree of agreement in the data geometry. Indeed, we found that RF resulted in uniformly smaller Procrustes distances from the embedding of the original data than did PCA at compression ratios between 5 and 100 (Fig.~\ref{fig:VaziriMouse}d). Moreover, the gap between PCA and RF was much starker in the LLE distances, indicating that even slight gains in the isometry constant can have profound impacts to data geometry preservation when it comes to more complex analyses. Altogether, both the isometry constants and Procrustes distances indicate that randomized filtering better preserves complex manifold structure than PCA.

\subsection*{Vortex dynamics are decodable from RF embeddings}

A second example of high-dimensional, complex dynamical data arises in the study of nonlinear fluid flow behavior such as turbulence---a fundamental phenomenon in fluid mechanics which can be notoriously difficult to model---via optical measurements of the vorticity fields.
Vorticity quantifies the local tendency for rotational movement in a fluid and, for the case of two dimensional incompressible flow, may be described as a scalar $\omega$ field that obeys the following system of partial differential equations\footnote{The function $\omega(x, y, t)$ is the scalar vorticity field, $\psi(x, y, t)$ is the stream function, $\gamma$ is the kinematic viscosity, $f(x,y)$ is a forcing function, $\nabla^\perp = (-\partial_y, \partial_x)^T$ is the perpendicular gradient operator, and $\Delta = \partial_{xx} + \partial_{yy}$ is the Laplace operator.}~\cite{Majda2002VorticityIncompressibleFlow}
\begin{alignat}{1}
\label{eqn:vorticity}
\partial_t \omega
+
\ps{ \nabla^\perp \psi} \cdot \nabla \omega
&=
\gamma \Delta \omega + f \nonumber \\
\Delta \psi &= \omega.
\end{alignat}

\begin{figure}[t]
\centering
\includegraphics{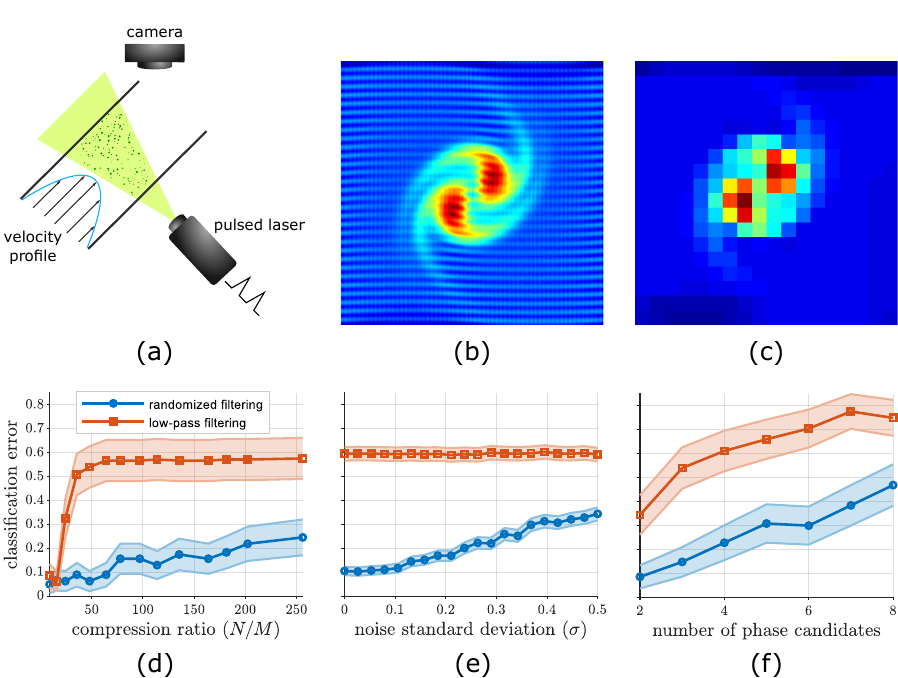}
\caption{
  {\bf (a)} Vorticity is a quantity that describes the local tendency of a fluid to rotate, and can be measured experimentally using, e.g., particle image velocimetry as shown.
  {\bf (b)} Example frame of the solution to the vorticity Equations~\eqref{eqn:vorticity}.
  {\bf (c) } Example dimensionality reduction by a factor of $15$ in each dimension via LPF and downsampling.
  As expected, the high-frequency content has vanished.
  {\bf (d-f) } Performance in a phase classification task on solutions to the vorticity equations. 
   The task is to classify which phase of the forcing function was used to generate the solution based on compressed measurements of the solution.
   {\bf (d) } RF achieves a low classification error rate even after 2 to 3 orders of magnitude of compression.
   {\bf (e) } RF produces superior phase classification performance in noisy conditions, whereas information necessary for classification is completely lost under low-pass filtering.
   {\bf (f) } RF allows for correct classification more often than LPF even when the problem is made more difficult by drawing phases from a larger set of candidates.
}
\label{fig:vorticity}
\end{figure}

Inspired by the simulations presented in~\cite{Schaeffer2013SparseDynamicsPartial}, we consider two vortex patches interacting under the influence of a high-frequency forcing function $f$ defined as
\begin{equation}
\label{eqn:forcing}
f(x,y) = 0.075 \frac{\sin(64x + \phi) + \sin(32 y + \phi)}
                {1 + 0.25(\cos(128x + \phi) + \cos(64y + \phi))}.
\end{equation}
We generated solutions to Equation~\eqref{eqn:vorticity} using a forcing function described by Equation~\eqref{eqn:forcing} with an unknown phase $\phi$ chosen uniformly at random from a set of phase candidates $\cb{\phi_1, \ldots, \phi_P}$. We then performed classification to determine which phase generated the observed vorticity field.
This classification simply selects the phase that minimizes the distance between the vorticity field at a given time and the forcing functions generated using each possible phase,
\begin{gather}
\est{\phi} = \argmin_{\phi_i}
\norm{\v{\Phi}\omega - \v{\Phi}f_{\phi_i}}_2^2,    
\end{gather}
where $f_{\phi_i}(x,y)$ is the forcing function produced by phase $\phi_i$ (see supplemental text).
The ability to capture the high frequency content of the forcing term (Fig.~\ref{fig:vorticity}b) allows RF to produce higher-accuracy phase estimates than equivalent dimensionality reduction using LPF (Fig.~\ref{fig:vorticity}c). Specifically, for compression ratios above $\approx 10$X, RF achieves a consistently good performance of $<0.2$ classification error, while LPF quickly rises to a classification error of $>0.5$ as early as 40X compression. Similar to the calcium imaging example (Fig.~\ref{fig:ca_sim}), RF remained stable both with respect to the noise levels (Fig.~\ref{fig:vorticity}e) and task complexity (Fig.~\ref{fig:vorticity}f).

\section{Discussion}

We present here randomized filtering (RF): a general, computationally efficient, and theoretically sound approach to dimensionality reduction. Our focus is on the ability to quickly reduce the dimension of large-scale streaming data at the very first stages of processing while maintaining critical information about the the data stream.
We demonstrate the benefits and performance of RF on several
applications, including neuroimaging, high-dimensional neural population activity, and vorticity estimation. Taken together, these applications demonstrate of the power of universal dimensionality reduction with RF.
Despite the vastly different datasets, the dimensionality reduction procedure used was identical.
It required no knowledge of either the data features or the post-compression task, making it extremely useful in discovery-oriented experiments where the characteristics of interest are yet unknown.

A key conceptual advance is the focus on preserving the low-dimensional data manifold. Compression always seeks to preserve specific ``important'' qualities. For example, image compression techniques such as JPEG~\cite{hudson2018jpeg} and JPEG2000~\cite{marcellin2000overview} aim to conserve the quality of the images as assessed by human visual perception. For many other datasets, however, it is unclear \emph{a priori} what the key factors to maintain are. The manifold assumption provides such a general quality that is well established and can be assessed quantitatively, as we show with the isometry calculations and additional analyses in classification, event detection, and manifold esimtation.    

One critical parameter in using any dimensionality reduction technique is the size of the embedding dimension---i.e., the compression ratio. We note that when using RF in a given application for the first time, choosing the single compression ratio parameter will require far less data compared to data-dependent methods which require full training to characterize the underlying structure.
Furthermore, this initial tuning need only be carried out once, after which the resulting value may be used in a variety of experimental conditions if the same underlying manifold model applies.
As RF becomes more widely used, the choice of compression ratio may become largely experience-driven.

We have explored the potential of RF by evaluating its performance on practical filtering tasks, but its applicability is in no way limited to this use case.
As RF preserves the manifold geometry underlying the observations, it is compatible with other techniques that depend on this structure, such as regression, classification and manifold learning approaches~\cite{Tenenbaum2000ISOMAP}, including Locally Linear Embedding~\cite{Roweis2000LLE}, which we demonstrate in our neural activity analysis example. In fact, in more general matrix factorization and time trace analysis~\cite{charles2022graft,mishne2019learning}, random projections are already being used to speed up computation~\cite{berlanga2025fast}, and RF serves as a critical catalyst to further speed up data analysis as the size of these datasets continues to grow. 
In imaging applications, RF could could even be implemented optically via a lens to calculate the DFT, removing the computational cost completely.

The development of new dimensionality reduction tools is essential in ensuring that data abundance is a blessing and not a curse.
Here we have demonstrated a simple yet powerful dimensionality reduction tool with wide applicability.
The shift from deterministic to randomized schemes represents a fundamental change in the way we think about dimensionality reduction, but it may be the only way to stay afloat as data volumes continue to surge. 


\section{Methods}
\subsubsection*{Preservation of inner products}
In this section, we introduce a result on the preservation of inner products which serves as theoretical justification for the approximation of the uncompressed estimator in the neuroimaging simulations.
Specifically, under the assumption that $\v{\Phi}$ is a stable embedding of $\man \cup -\man$ (i.e., that $\v{\Phi}$ also stably embeds the manifold's reflection about the origin), we develop a bound for the deviation of inner products computed in the reduced space versus the original space.
We note that the bound of \cite[Theorem 4]{Davenport2010CompressiveSP} cannot be applied directly due to its implicit assumption that $\v{\Phi}$ is also a stable embedding of the unit vectors in the direction of each point on $\man$.
Instead, we have the following theorem.

\setcounter{theorem}{1}
\begin{theorem}
\label{thm:ip_pres}
Suppose that $\ell, s \in \man$ and that $\Phi$ is a $\delta$-stable embedding of $(\man \cup -\man)$, then
\[
\abs{\ip{\Phi \ell}{\Phi s} - \ip{\ell}{s}} \leq \frac{\delta}{2}(\norm{\ell}_2^2 + \norm{s}_2^2).
\]
\end{theorem}
\begin{proof}
Since $\Phi$ is a stable embedding of $(\man \cup -\man)$, we have
\[
1-\delta \leq \frac{\norm{\Phi \ell \pm \Phi s}^2}{\norm{\ell \pm s}^2} \leq 1+\delta,
\]
or equivalently (by expanding the norm in the denominator)
\[
1-\delta \leq \frac{\norm{\Phi \ell \pm \Phi s}^2}{\norm{\ell}^2 + \norm{s}^2 \pm 2\ip{\ell}{s}} \leq 1+\delta.
\]
By the parallelogram law,
\begin{alignat*}{1}
\ip{\Phi \ell}{\Phi s} = &\enspace
\frac{\norm{\Phi \ell + \Phi s}^2 - \norm{\Phi \ell - \Phi s}^2}{4}
\\ \leq &\enspace
\frac{(1+\delta)(\norm{\ell}^2 + \norm{s}^2 + 2\ip{\ell}{s}) - (1-\delta)(\norm{\ell}^2 + \norm{s}^2 - 2\ip{\ell}{s})}{4}
\\ = &\enspace
\ip{\ell}{s} + \frac{\delta}{2}(\norm{\ell}^2 + \norm{s}^2),
\end{alignat*}
Similarly, we can show that
\[
\ip{\Phi \ell}{\Phi s} \geq \ip{\ell}{s} - \frac{\delta}{2}(\norm{\ell}^2 + \norm{s}^2).
\]
Taken together, these two inequalities yield the desired result
\[
-\frac{\delta}{2} (\norm{\ell}^2 + \norm{s}^2) \leq \ip{\Phi \ell}{\Phi s} - \ip{\ell}{s} \leq \frac{\delta}{2} (\norm{\ell}^2 + \norm{s}^2).
\]
\end{proof}

\begin{remark}
If we further assume that $x/\norm{x}_2 \in \man \enspace \forall x \in \man$, then we recover the result in \cite[Theorem 4]{Davenport2010CompressiveSP} which deals mostly with sparse signal models rather than the manifold models considered here.
\end{remark}

\subsubsection*{Validation of scaling behavior via sine manifold}

RF is a stable embedding of the input manifold when $M$ is larger than a quantity which scales \emph{logarithmically} with the ambient dimension $N$.
We verify this fact by computing the value of $M$ required to achieve a fixed isometry constant $\delta$.
As a point of reference, we set $\delta=1/3$ which is the threshold used in related manifold recovery results \cite{Shah2011IterativeProjectionsSignal}.
The input data for this simulation is the sine manifold described by \eqref{eqn:sine_manifold}.
For values of $N < 1000$, the scaling is logarithmic as expected.
For larger values of $N$, the required dimension of the reduced space increases even more slowly.

\begin{figure}
\centering
\includegraphics{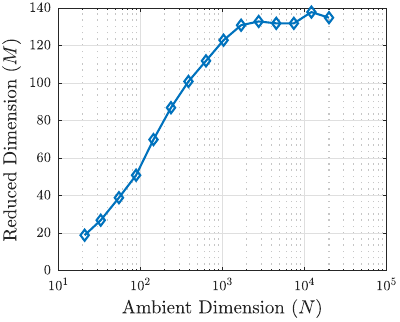}
\caption{
  Scaling behavior of RF with the sine manifold for a fixed isometry constant $\delta=1/3$.
}
\end{figure}

\subsubsection*{Estimation with complex-valued measurements: real-valued noise}

Let $\v{X} \sim \normal(\v{s}, \sigma^2 \v{I}_N)$ and $\v{Z} = \v{\Phi X}$ where $\v{\Phi} \in \complex^{M \times N}$.
Suppose we wish to describe the density function associated with $\v{Z}$ in order to compute a maximum likelihood estimate (MLE) of one of the model parameters.
Since the mean and covariance matrix of $\v{Z}$ are given by
\begin{equation*}
\Exp{\v{Z}} = \Exp{\v{\Phi X}} = \v{\Phi}\Exp{\v{X}} = \v{\Phi s},
\end{equation*}
and
\begin{alignat}{1}
\Cov{\v{Z}}{\v{Z}} &\enspace =
\Exp{(\v{Z} - \Exp{\v{Z}})(\v{Z}-\Exp{\v{Z}})^*} \nonumber \\
&\enspace =
\Exp{(\v{\Phi X} - \Exp{\v{\Phi X}})(\v{\Phi X}-\Exp{\v{\Phi X}})^*} \nonumber \\
&\enspace =
\v{\Phi}\Exp{(\v{X} - \Exp{\v{X}})(\v{X}-\Exp{\v{X}})^*}\v{\Phi}^* \nonumber \\
&\enspace =
\v{\Phi}\Cov{\v{X}}{\v{X}}\v{\Phi}^* \nonumber \\
&\enspace =
\v{\Phi}\v{\Phi}^*,
\end{alignat}
respectively, it may be tempting to express the distribution using the well-known formula for the multivariate Gaussian, i.e.,
\begin{equation}
\label{eqn:complex_gaussian_pdf}
f_{\v{Z}}(\v{z} \given \v{s})
=
(2\pi)^{-n/2} \det(\v{\Phi\Phi}^*)^{-1/2}
\exp\cb{-\frac{1}{2\sigma^2} (\v{z} - \v{\Phi s})^*(\v{\Phi \Phi}^*)^{-1} (\v{z} - \v{\Phi s})},
\end{equation}
where $\v{\Phi}^*$ denotes the conjugate transpose of the matrix $\v{\Phi}$.
Unfortunately, the mean and covariance matrix alone are insufficient to fully specify the distribution of $\v{Z}$.
To see this, it is useful to instead consider the real-valued random vector produced by the canonical isomorphism from $\complex^m \to \reals^{2m}$ given by
\begin{equation*}
\v{a} + i\v{b} \mapsto \begin{bmatrix} \v{a} \\ \v{b} \end{bmatrix}.
\end{equation*}
In our problem, we can generate this vector via the measurement matrix
\begin{equation*}
\v{A} \coloneqq \begin{bmatrix} \v{\Phi}_R \\ \v{\Phi}_I \end{bmatrix},
\end{equation*}
where $\v{\Phi}_R = \Re\cb{\v{\Phi}}$ and $\v{\Phi}_I = \Im\cb{\v{\Phi}}$ denote the real and imaginary components of $\v{\Phi}$.
Denoting the resulting measurements by $\v{Y} = \v{A X}$,
we know from the analysis of real Gaussian vectors that the distribution of $\v{Y}$ is determined by the mean $\v{As}$ and covariance matrix $\v{AA}^T$.
The likelihood function is then given by\footnote{We assume here that $\v{AA}^T$ is non-singular. In the singular case, one may consider a subset of the entires in $\v{Y}$ such that the corresponding covariance matrix is non-singular. Alternatively, we may replace the determinant and inverse with the pseudo-determinant and pseudo-inverse respectively.}
\begin{equation*}
f_{\v{Y}}(\v{y} \given \v{s})
=
(2\pi)^{-n/2} \det(\v{AA}^T)^{-1/2}
\exp\cb{-\frac{1}{2\sigma^2} (\v{y} - \v{A s})^T(\v{AA}^T)^{-1} (\v{y}_t - \v{A s})}.
\end{equation*}

To understand why we cannot use the likelihood function \eqref{eqn:complex_gaussian_pdf}, we will argue that the mean and covariance matrix are insufficient to describe the distribution of $\v{Y}$.
To begin, we note that the covariance matrix $\v{AA}^T$ may be written in block form as
\begin{equation*}
\v{AA}^T =
\begin{bmatrix}
\v{\Phi}_R \v{\Phi}_R^T & \v{\Phi}_R \v{\Phi}_I^T \\
\v{\Phi}_I \v{\Phi}_R^T & \v{\Phi}_I \v{\Phi}_I^T.
\end{bmatrix}
\end{equation*}
In contrast, the covariance matrix associated with $\v{y}$ is given by
\begin{equation*}
\v{\Phi \Phi}^* =
(\v{\Phi}_R \v{\Phi}_R^T + \v{\Phi}_I \v{\Phi}_I^T)
+ i(\v{\Phi}_I \v{\Phi}_R^T - \v{\Phi}_R \v{\Phi}_I^T).
\end{equation*}
Since $\v{\Phi \Phi}^*$ contains only sums of the blocks of $\v{AA}^T$, it is impossible to infer the covariance matrix of $\v{Y}$---and hence the distribution of $\v{Z}$---from the covariance matrix of $\v{Z}$ alone.

To specify the distribution of $\v{Z}$, we require one additional quantity called the \emph{pseudo-covariance} matrix defined by \cite{Tse2005FundamentalsWirelessCommunication}
\begin{equation}
\label{eqn:psuedo_cov}
\Exp{\v{ZZ}^T}
= \v{\Phi\Phi}^T =
(\v{\Phi}_R \v{\Phi}_R^T - \v{\Phi}_I \v{\Phi}_I^T)
+ i(\v{\Phi}_I \v{\Phi}_R^T + \v{\Phi}_R \v{\Phi}_I^T).
\end{equation}
The covariance and pseudo-covariance matrices of $\v{Z}$ be used to compute the full covariance matrix of $\v{Y}$.
Thus the mean, covariance, and pseudo-covariance of $\v{Z}$ fully specify its distribution.

If the pseudo-covariance matrix of the noise is equal to zero, then it follows a special type of complex Gaussian distribution called the \emph{circularly symmetric Gaussian} \cite{Lapidoth2009FoundationDigitalCommunication}.
In this case, \eqref{eqn:psuedo_cov} yields
\begin{equation*}
\v{\Phi}_R \v{\Phi}_R^T = \v{\Phi}_I \v{\Phi}_I^T,
\quad
\v{\Phi}_I \v{\Phi}_R^T = -\v{\Phi}_R \v{\Phi}_I^T.
\end{equation*}
The covariance matrix of $\v{Y}$ then becomes
\begin{equation*}
\v{AA}^T =
\begin{bmatrix}
\v{\Phi}_R \v{\Phi}_R^T & \v{\Phi}_R \v{\Phi}_I^T \\
-\v{\Phi}_R \v{\Phi}_I^T & \v{\Phi}_R \v{\Phi}_R^T
\end{bmatrix},
\end{equation*}
and each block may be computed using the covariance matrix of $\v{Z}$ which is given by
\begin{equation*}
\v{\Phi \Phi}^* = 2\v{\Phi}_R \v{\Phi}_R^T - 2i \v{\Phi}_R \v{\Phi}_I^T.
\end{equation*}
In other words, the mean and covariance matrix are sufficient to determine the distribution of circularly symmetric Gaussian vectors.
Furthermore, it may be shown that the corresponding density function is of the form in \eqref{eqn:complex_gaussian_pdf}.
In the setting of this paper, the noise is not circularly symmetric as presented in the main text, so we carry out analysis using $\v{Y}$.
However, assuming a circularly symmetric complex noise model as an approximation to the real noise model simplifies analysis and yields connections to estimators in the uncompressed space, so we also consider this noise model in subsequent sections.

\subsubsection*{Estimation with complex-valued measurements: complex-valued noise}

We now consider a variation on the problem in the previous section where we instead assume complex-valued noise.
We denote by $\complexnormal\ps{\v{0}, \v{K}}$ the distribution of of a circularly symmetric Gaussian random vector with covariance matrix $\v{K}$.
Our signal model may then be expressed as $\v{X} \sim \v{s} + \v{\epsilon}$ where $\v{\epsilon} \sim \complexnormal\ps{0, \sigma^2 \v{I}_N}$.
As before, we observe data through a complex measurement operator $\v{\Phi} \in \complex^{M \times N}$ via $\v{Z} = \v{\Phi X}$ and wish to estimate a parameter of the model.
As in the previous section, the covariance matrix of $\v{Z}$ is given by $\v{\Phi \Phi}^*$, but in this case, since the noise is circularly symmetric, $\v{s}$ and the covariance matrix are sufficient to fully specify the density function:
\begin{equation}
\label{eqn:complex_normal_pdf}
f_{\v{Z}}(\v{z} \given \v{s})
=
\frac{1}{\pi^n \sigma^2 \det(\v{\Phi\Phi}^*)}
\exp\cb{-\frac{1}{\sigma^2} (\v{z} - \v{\Phi s})^*(\v{\Phi \Phi}^*)^{-1} (\v{z}_t - \v{\Phi s})}.
\end{equation}
We will see in subsequent sections that estimators simplify especially well under this noise model and the randomized filtering matrix $\v{\Phi}$.

\subsubsection*{Application: time series estimation from reduced measurements}
We now derive the estimator given in \eqref{eqn:param_est_original}.
Suppose $\v{s}$ is a spatial cell profile represented by a vectorized binary mask (i.e., $\v{s}[i] = 1$ if pixel $i$ contains the cell of interest and 0 otherwise).
Let $\v{x}_t = \mu_t \v{s} + \sigma \v{\epsilon}$ represent the noisy image at time $t$ whose mean is the cell template modulated by a time varying parameter $\mu_t$ which we wish to estimate.
In the original space, the distribution function of $\v{y}_t$ is then given by
\begin{equation*}
f_{\v{X}}(x_t \given \mu_t, \v{s}) =
(2\pi)^{-n/2}
\prod_{i} \exp\cb{-\frac{1}{2\sigma^2}(\v{x}_t - \mu_t \v{s})^T(\v{x}_t - \mu_t \v{s})},
\end{equation*}
Setting the derivative with respect to $\mu_t$ of the log likelihood equal to zero yields
\begin{equation*}
\widehat{\mu}_t
= \frac{\v{x}^T \v{s}}{\norm{\v{s}}^2},
\end{equation*}
which is the estimator \eqref{eqn:param_est_original}.

Now suppose we are given measurements in the reduced space given by $\v{z}_t = \v{\Phi x}_t$.
Defining the equivalent measurement matrix $\v{A} = [\v{A}_R; \v{A}_I]$ and letting $\v{y}_t = \v{A x}_t$, we have the distribution function given by
\begin{equation*}
f_{\v{Y}}(\v{y}_t \given \mu_t, \v{s})
= C
\exp\cb{-\frac{1}{2\sigma^2} (\v{y}_t - \mu_t\v{A s})^T(\v{AA}^T)^{-1} (\v{y}_t - \mu_t\v{A s})},
\end{equation*}
where $C = (2\pi)^{-n/2} \det(\v{AA}^T)^{-1/2}$.
Setting the derivative log likelihood equal to zero yields the following estimator:
\begin{equation}
\label{eqn:true_mle}
\est{\mu_t} = \frac{\v{y}_t^T (\v{AA}^T)^{-1} \v{As}}{(\v{As})^T (\v{AA}^T)^{-1} \v{As}}.
\end{equation}
Thus, the estimator is an inner product between the measurements and the cell template in the reduced space after whitening by the matrix $(\v{AA}^T)^{-1}$.

While \eqref{eqn:true_mle} is the technically correct MLE for $\mu_t$, it is also instructive to consider the estimator under a circularly symmetric complex noise model.
In this case, the compressed measurements $\v{z}$ are then distributed as
\begin{equation*}
\v{z}_t \sim \complexnormal\ps{\mu_t \v{\Phi} \v{s}, \sigma^2 \v{\Phi\Phi}^*},
\end{equation*}
and by \eqref{eqn:complex_normal_pdf} we have the distribution function
\begin{equation*}
f_{\v{Z}}(\v{z}_t \given \mu_t, \v{s})
=
\frac{1}{\pi^n \det(\v{\Phi\Phi}^*)}
\exp\cb{-\frac{1}{\sigma^2} (\v{z}_t - \mu_t\v{\Phi s})^*(\v{\Phi \Phi^*})^{-1} (\v{z}_t - \mu_t\v{\Phi s})}.
\end{equation*}
Maximizing the likelihood is equivalent to minimizing the quadratic form in the exponential which we do by computing its derivative:
\begin{alignat}{1}
&\frac{\partial Q}{\partial \mu_t} =
\frac{\partial}{\partial \mu_t}
\ps{(\v{z}_t - \mu_t\v{\Phi s})^*(\v{\Phi \Phi^*})^{-1} (\v{z}_t - \mu_t\v{\Phi s})}
\nonumber \\=&
\frac{\partial}{\partial \mu_t} \ps{
  \v{z}_t^*(\v{\Phi \Phi}^*)^{-1} \v{z}_t
  - \mu_t\v{z}_t^*(\v{\Phi \Phi}^*)^{-1} \v{\Phi s}
  - \mu_t(\v{\Phi s})^*(\v{\Phi \Phi}^*)^{-1} \v{z}_t
  + \mu_t^2 (\v{\Phi s})^*(\v{\Phi \Phi}^*)^{-1} \v{\Phi s}
  }
\nonumber \\=&
\frac{\partial}{\partial \mu_t} \ps{
  \v{z}_t^*(\v{\Phi \Phi}^*)^{-1} \v{z}_t
  - \mu_t\v{z}_t^*(\v{\Phi \Phi}^*)^{-1} \v{\Phi s}
  - \mu_t(\v{z}_t^*(\v{\Phi \Phi}^*)^{-1} \v{\Phi s})^*
  + \mu_t^2 (\v{\Phi s})^*(\v{\Phi \Phi}^*)^{-1} \v{\Phi s}
  }
\nonumber \\=&
\frac{\partial}{\partial \mu_t} \ps{
  \v{z}_t^*(\v{\Phi \Phi}^*)^{-1} \v{z}_t
  - 2\mu_t\Re\cb{\v{z}_t^*(\v{\Phi \Phi}^*)^{-1} \v{\Phi s}}
  + \mu_t^2 (\v{\Phi s})^*(\v{\Phi \Phi}^*)^{-1} \v{\Phi s}
  }
\nonumber \\=&
  - 2\Re\cb{\v{z}_t^*(\v{\Phi \Phi}^*)^{-1} \v{\Phi s}}
  + 2\mu_t (\v{\Phi s})^*(\v{\Phi \Phi}^*)^{-1} \v{\Phi s}
\end{alignat}
As before, we set the derivative with respect to $\mu_t$ of the log likelihood equal to zero which yields the MLE for a general sensing matrix $\v{\Phi}$:
\begin{equation*}
\widehat{\mu}_t =
\frac
  {\Re\cb{\v{z}_t^* (\v{\Phi \Phi^*})^{-1} \v{\Phi s}}}
  {(\v{\Phi s})^* (\v{\Phi \Phi^*})^{-1} \v{\Phi s}}.
\end{equation*}

In the RF setting, the estimator simplifies significantly.
Note that the construction of $\v{\Phi}$ for randomized filtering may be expressed as $\v{\Phi} = \v{SFD}$ where $\v{S}$ is the matrix which selects a subset of the rows in the matrix it multiplies, $\v{F}$ is the DFT matrix, and $\v{D}$ contains a Rademacher sequence on its diagonal.
Then we have
\begin{equation*}
\v{\Phi\Phi}^* = \v{SFD}\v{D}\v{F}^* \v{S}^T = \v{I}_M,
\end{equation*}
where $\v{I}_M$ is the $M \times M$ identity matrix.
Thus, the estimator simplifies to
\[
\widehat{\mu}_t = \frac{\Re\cb{\v{z}_t^* \v{\Phi s}}}{\norm{\v{\Phi s}}^2},
\]
which is the estimator \eqref{eqn:ip}.
Alternatively, in light of Theorem~\ref{thm:ip_pres}, we may view the estimator in the compressed space as an approximation of the one in the original space.

\subsubsection*{Application: classification from reduced measurements}

Suppose we measure noisy observations of an image with mean equal to an unknown element $\v{s}^{\star}$ from a set of known candidates $S$:
\begin{equation*}
\v{x} \sim \normal({\v{s}^{\star}, \sigma^2 \v{I}_N}), \quad \v{s}^{\star} \in S.
\end{equation*}
The density function is given by
\begin{equation*}
f_{\v{X}}(\v{x} \given \v{s}_i)
=
(2\pi)^{-n/2}
\exp\cb{-\frac{1}{2\sigma^2} (\v{x} - \v{s}_i)^T(\v{x} - \v{s}_i)},
\end{equation*}
which we maximize with respect to $i$ yielding the classifier:
\begin{equation*}
\est{\v{s}} = \argmin_{\v{s}_i \in S} \norm{\v{x} - \v{s}_i}^2.
\end{equation*}

Now suppose we observe reduced measurements via the RF operator $\v{\Phi}$.
As before, we consider the equivalent measurements through the corresponding operator $\v{A}$ which yields the likelihood function
\begin{equation*}
f_{\v{Y}}(\v{y} \given \v{s}_i)
=
(2\pi)^{-n/2} \det(\v{AA}^T)
\exp\cb{-\frac{1}{2\sigma^2} (\v{y} - \v{A}\v{s}_i)^T(\v{AA}^T)^{-1}(\v{y} - \v{A}\v{s}_i)},
\end{equation*}
so the corresponding classifier is
\begin{equation*}
\est{\v{s}} = \argmin_{\v{s}_i \in S}(\v{y} - \v{A}\v{s}_i)^T(\v{AA}^T)^{-1}(\v{y} - \v{A}\v{s}_i).
\end{equation*}

Finally, if we approximate the noise as circularly symmetric and note that $\v{\Phi \Phi}^* = \v{I}_M$, the likelihood function may be expressed as
\begin{equation*}
f_{\v{Z}}(\v{z} \given \v{s}_i)
=
(2\pi)^{-n/2}
\exp\cb{-\frac{1}{2\sigma^2} (\v{z} - \v{\Phi s}_i)^*(\v{z} - \v{\Phi s}_i)},
\end{equation*}
which yields the classifier
\begin{equation*}
\est{\v{s}} = \argmin_{\v{s}_i \in S} \norm{\v{z} - \v{\Phi s}_i}^2.
\end{equation*}
Again, we may view this as an approximation to the classifier in the original space which has error bounded by the isometry constant $\delta$ by virtue of the fact that $\v{\Phi}$ is a stable embedding.

\subsubsection*{Isometry constant estimation}
Here we outline the procedure for estimating the isometry constant $\delta$ in Figure~\ref{fig:rip}.
For a given dataset $\v{D} \in \reals^{N\times K}$ with $K$ samples of dimension $N$, we compute all pairwise distances in the original space:
\[
d_{\text{o}}(i, j) = \norm{\v{x}_i - \v{x}_j}_2, \quad i < j \leq K,
\]
and in the reduced space
\[
d_{\text{r}}(i, j) = \norm{\v{y}_i - \v{y}_j}_2, \quad i < j \leq K,
\]
where $\v{Y} = \v{\Phi D}$.
Next let $Q(i,j)$ denote the ratio of pairwise distances in the reduced space to those in the original space, i.e.,
\[
Q(i,j) = \frac{d_{\text{o}}(i,j)}{d_{\text{r}}(i,j)},
\]
and define
\[
Q_{\text{min}} = \min_{ij} Q(i,j), \quad
Q_{\text{max}} = \max_{ij} Q(i,j), \quad
Q_{\text{mean}} = \frac{2}{K(K-1)} \sum_{i<j} Q(i,j).
\]
We estimate the lower and upper isometry constants as
\[
\delta_{\text{lower}} = 1-\frac{Q_{\text{min}}}{Q_{\text{mean}}},
\quad \delta_{\text{upper}} = \frac{Q_{\text{max}}}{Q_{\text{mean}}}-1,
\]
where we normalize by the mean ratio to account for any constant scaling caused by $\v{\Phi}$.
Finally, we take $\delta$ to be the worse of the two values above:
\[
\delta = \max(\delta_{\text{lower}}, \delta_{\text{upper}}).
\]

\subsection*{Datasets}
\subsubsection*{Sine manifold}
The sine manifold, also called the complex exponential curve \cite{Eftekhari2015NewAnalysisManifold}, is a one dimensional manifold in $\complex^N$ defined by
\begin{equation}
\label{eqn:sine_manifold}
\v{y}(t) =
\begin{bmatrix}
e^{-i2\pi f_C t} \\
e^{-i2\pi (f_C-1) t} \\
\vdots \\
e^{i2\pi (f_C-1) t} \\
e^{i2\pi f_C t}
\end{bmatrix},
\end{equation}
where $f_C$ is an integer that defines the dimension of the ambient space via $N=2f_C + 1$.
Because the sine manifold can be easily embedded in any arbitrary ambient dimension, and because bounds exist for its manifold properties (e.g., its reach), the sine manifold is an informative toy example in the analysis of manifold embeddings \cite{Eftekhari2015NewAnalysisManifold}.
The results in Figure~\ref{fig:rip} are for a sine manifold with $f_C = 5,000$ resulting in an ambient dimension of $N=10,001$.

\subsubsection*{Cardiac simulation}

Mathematical modeling of the heart plays a key role in the development of treatments for patients with cardiovascular disease.
Figure~\ref{fig:rip} shows the isometry constant for RF applied to data from a modified O'Hara-Vir\'{a}g-Varr\'{o}-Rudy (OVVR) model \cite{Elshrif2014QuantitativeComparisonBehavior}.
The model we use incorporates real clinical data and faithfully reproduces key features of action potentials (e.g. duration, amplitude and resting membrane potential) for patients with heart failure.
The data used here represent the voltage on a three dimensional grid consisting of $N=4,129$ points over $K=12,500$ time samples.

\subsubsection*{Calcium imaging (simulated)}

Calcium imaging is a staple neural recording technology in modern neuroscience.
We generate data which simulates some of the key features of wide field calcium imaging recordings while simultaneously allowing full control over parameters of interest.
Our data generation procedure consists of three components: creation of cell spatial profiles, creation of temporal events, and synthesis of the final dataset.
Cell spatial profiles are generated sequentially by first centering a two dimensional Gaussian with radius $r$ at a location chosen uniformly at random.
The radius $r$ follows a Gaussian distribution with a mean of one pixel and a standard deviation of 0.1.
The next cell is generated in the same way, but with probability $p_o$ the location is chosen such that the cell overlaps with the previous one.
With probability $1-p_o$, the cell location is again chosen uniformly at random.
This process repeats until $N_c$ cells have been placed and the resulting profiles are vectorized and stored as the columns of the $N \times N_c$ matrix $\v{S}$.
The results in Figure~\ref{fig:ca_sim} use $N_c=512$ cells spread over a $256 \times 256$ pixel field of view.

To generate the temporal event profiles, we perform the following procedure for each cell:
first, generate a spike train in which the time between spikes follows the exponential distribution (i.e., a Poisson process), but with an additional constraint on the minimum time between spikes.
This modification to the standard Poisson process simulates the refractory period during which a cell which has just fired is unable to fire again for a specified duration of time.
The rate parameter for each cell is equal to one and the refractory period is 0.3.
Next, we convolve the resulting spike train with a Gaussian bump to simulate the diffuse temporal profile commonly seen in calcium imaging recordings.
The resulting temporal profile for each cell is stored in the $N_t \times N_c$ matrix $\v{T}$.

Finally, data matrix is constructed by modulating each spatial profile by its corresponding temporal profile and adding noise via $\v{D} = \v{ST}^T + \sigma \v{\epsilon}$.

\subsubsection*{Calcium imaging from mouse}

Manley \textit{et al.}~\cite{Manley2024,VaziriMouseData} generated a set of open source large-scale neural datasets using light beads calcium imaging. In this paper, we analyzed a single-region recording from the right cortical hemisphere of a mouse over the course of 1 hour (6519 neurons, 34600 time points, 9.6 Hz volume rate). This work from the Vaziri lab was notable for its scale: up to one million neurons recorded simultaneously. Furthermore, Manley \textit{et al.} observed that much of the data required a high-dimensional representation. When analyzing a small, single-region subset of this data with RF and PCA, we found that while PCA causes significantly distortion (in terms of isometry constant and Procrustes distance) even at weak compression, thus requiring a high-dimensional representation, RF manages to prevent manifold distortion even at stronger compression levels. Future work may determine whether these RF findings scale to hundreds of thousands of neurons, or whether the larger datasets are intrinsically high-dimensional, as claimed.

\subsubsection*{Vorticity equations}

The vorticity quantifies a fluid's localized tendency to rotate.
The equations describing vorticity can be derived from the laws of conservation of mass and conservation of momentum, or from the Navier-Stokes equations.
Here we consider shallow fluid flow which is well approximated by setting up the problem in two dimensions instead of three.
For the results in Figure~\ref{fig:vorticity}, we specify periodic boundary conditions and an initial condition with two vortex patches next to one another defined by
\begin{equation*}
\omega_0(x,y) = e^{-17.5(x-0.45)^2 - 0.7 y^2} + e^{-17.5(x+0.45)^2 - 0.7y^2}.
\end{equation*}
We solve the vorticity equations \eqref{eqn:vorticity} via spectral methods which are particularly well-suited for problems defined over rectangular domains with periodic boundary conditions, yielding fast solutions with accuracy beyond all algebraic orders.

First, we rewrite the top equation in \eqref{eqn:vorticity} as
\begin{equation*}
\partial_t \omega = \gamma \Delta\omega + f - (\nabla^\perp\psi)\cdot \nabla \omega.
\end{equation*}
Computing the Fourier transform of the both sides with respect to $x$ and then $y$, we have
\begin{alignat}{1}
\ft{\partial_t \omega}
=&\enspace
  \ft{\gamma \Delta\omega + f - (\nabla^\perp\psi)\cdot \nabla \omega}
\nonumber \\ =& \enspace
  -\gamma \ft{\omega}(k_x^2 + k_y^2)
  + \ft{f (\nabla^\perp\psi)\cdot \nabla \omega},
\end{alignat}
where $k_x$ and $k_y$ are the wave numbers over $x$ and $y$ respectively.
Similarly, for the bottom equation of \eqref{eqn:vorticity} we have
\begin{equation*}
\ft{\omega} = \frac{\ft{\psi}}{k_x^2 + k_y^2}.
\end{equation*}
We then use MATLAB's ode45 ordinary differential equation solver to compute $\ft{\omega}$ and recover $\omega$ via the inverse transform.

\subsubsection*{Functional magnetic resonance imaging}

Functional magnetic resonance imaging measures the brain by using blood flow as proxy for neural activity.
Figure~\ref{fig:rip} shows the isometry constant for the publicly available Enhanced Nathan Kline Institute-Rockland Sample dataset \cite{Nooner2012NKIRocklandSampleModel}.
We use slices 30--33 which contain a total of 24,964 voxels over 900 samples in time.

\subsubsection*{Voltage sensitive dye}

Voltage sensitive dye (VSD) is an imaging technique which enables recordings of activations across multiple cortical columns for use in scientific applications.
Figure~\ref{fig:rip} shows the isometry constant for the dataset reported in \cite{Zheng2015AdaptiveShapingCortical} which uses VSD to record from the primary sensory cortex in anesthetized rats for the purpose of studying adaptation to stimuli in the vibrissa pathway.
The dataset contains recordings from multiple experimental trials for a total of 3,000 image samples containing 22,632 pixels per image.

\section*{Acknowledgments}

EY and ASC were supported in part by NSF CAREER award 2340338. NPB and CJR were supported by NSF grant number CCF-1409422. HLY and CJR were supported by NSF grant number CCF-0830456. CJR was supported by James S. McDonnell Foundation grant number 220020399. We would like to thank Dr. Linwei Wang's lab for sharing their cardiac simulation data, Dr. Garrett Stanley's lab for their voltage-sensitive dye data~\cite{Zheng2015AdaptiveShapingCortical}, and Dr. Alipasha Vaziri's lab for their open-source mouse data~\cite{VaziriMouseData}.

\section*{Data and Code Availability}

Code will be made available on GitHub upon publication; datasets are available as described in the publications referenced in Appendix Section ``Datasets.''

\renewcommand\refname{References and Notes}
\bibliography{refs_EvaEdits,notes}
\bibliographystyle{plain}

\end{document}